# Switching response and ionic hysteresis in organic electrochemical transistors

**Juan Bisquert,[1, 2]\* Baurzhan Ilyassov\*[3] and Nir Tessler[4]\***

[1] Instituto de Tecnología Química (Universitat Politècnica de València-Agencia Estatal Consejo Superior de Investigaciones Científicas), Av. dels Tarongers, 46022, València, Spain.

[2]Institute of Advanced Materials (INAM), Universitat Jaume I, 12006 Castelló, Spain

[3]Astana IT University, Mangilik El 55/11, EXPO C1, 010000 Astana, Kazakhstan

[4]Andrew & Erna Viterbi Department of Electrical and Computer Engineering, Technion, Haifa 32000, Israel

Email: bisquert@uji.es   nir@technion.ac.il

**Abstract**

Hysteresis in organic electrochemical transistors (OECT) is a basic effect in which the measured current depends on the voltage sweep direction and velocity. This phenomenon has an important impact on different aspects of the application of OECT, such as the switching time and the synaptic properties for neuromorphic applications. Here we address the combined ionic and electronic kinetic effects that cause the dominant hysteresis effects. We use a combination of tools consisting on basic analytical models, advanced 2D drift-diffusion simulation, and the experimental measurement of a Poly(3-hexylthiophene) (P3HT) OECT, working in an accumulation mode. We develop a general transmission line model considering drift electronic transport and ionic injection and diffusion from the electrolyte. We provide a basic classification of the transient response to a voltage pulse, according to the dominant ionic or electronic relaxation time, and the correspondent hysteresis effects of the transfer curves according to the general categories of inductive and capacitive hysteresis. These are basically related to the main control phenomenon, either the vertical diffusion of ions during doping and dedoping, or the equilibration of electronic current along the channel length.

## 1. Introduction

The realization of the point contact transistor in 1947 marked the beginning of the transistor era and the development of various transistor structures aiming to replace the vacuum-tube technology.[1] The demonstration of the first microprocessor in 1971 by Intel turned the metal oxide semiconductor field effect transistor (MOSFET) into the dominant transistor technology. Interestingly, before founding Intel, Gordon Moore extrapolated 4 years of data points to predict that the number of transistors per chip would double every 2 years.[2] This soon became the industry's target and what is known today as Moore's law. For many years, the law was followed by "simply" shrinking the size of the transistor. However, as the challenges accumulated, new transistor architectures were developed, and 3D integration became necessary. In parallel, a trend of improving performance not just by packing more transistors but also by performing the functions differently started evolving.[3]

The appearance of artificial intelligence and the notion of multi-level logic has accelerated the development of multi-level switches, the most famous of which is the memristor family.[4] Recent developments show that the good-old MOSFET transistor and its organic analogue[5] can also be turned into a multi-level switch by embracing electrochemistry as part of the transistor's toolbox.[6-8] Specifically, the electrochemical RAM (ECRAM) and the organic electrochemical transistor (OECT) use ionic and electronic conduction to establish new operation mechanisms, with the OECT also providing an interface to the biological world.[9] Recently, three terminal devices have been investigated as programmable resistors for brain-like computation. If the ion reservoir is left electrically open, the ions remain in the channel at the set conductivity, with non volatile property. This device class includes ENODe (Electrochemical Neuromorphic Device), and EIS (Electrochemical Ionic Synapse) configurations.[10-15]

It has been widely recognized that the dynamics of charge and ion transport exert a dominant influence in the operation of OECT.[16-18] In particular slow ionic motion often creates a kinetic limiting effect.[19,20] These characteristic dynamics have an important influence on the switching time of the OECT, and on the memory effects for neuromorphic applications. An important feature that has been observed in OECT is the hysteresis in the transfer curves.[16,21,22]

The analysis of transient behaviour of organic transistors has been developed by a standard distributed transmission line approach for both injection of electronic currents from drain and source, and injected ionic current from the electrolyte.[23-28] A particular realization is the Bernard and Malliaras model[29] that has been widely used for the characterization of transient response of the OECTs.[24,29-31] Recently we have developed a model of hysteresis in OECT[32] that consists on an extension of the Bernards and Malliaras model by coupling the diffusion of intercalated ions explicitly. The analytical model provides some useful distinctions of hysteresis effects provided that the density of ions can be considered nearly homogeneous in the film. Here we establish a rigorous general formulation of the model, and the suitable approximations that can be used to formulate the current response to different types of kinetic measurements, such as voltage scan at a constant rate, and sudden connection. Our method is to incorporate the diffusion of ions in the vertical direction in a transmission line formalism, so that we can determine different kinetic times constants that govern the charging and discharging of the channel towards the steady-state situation, and the correspondent hysteresis phenomena.

Below we apply the model investigation of the main hysteresis effects in both accumulation and depletion semiconductors.[27] We also explore the realistic operation and hysteresis effects using a 2-D drift-diffusion simulation model, that takes into account the ionic distribution and diffusion both in the organic film and the electrolyte.[33-37] Furthermore we show some characteristic experimental results of a Poly(3-hexylthiophene) (P3HT) OECT, working in an accumulation mode. These results provide a consistent picture of the interaction of ionic and electronic affects that govern the major hysteresis effects observed in OECT. The simulation allows one to explore different conditions of inhomogeneity, and a new mechanism is identified where the channel current follows the build-up of the anion density at the solution/semiconductor interface.

## 2. Model
### 2.1. Geometry of the model and carrier distribution

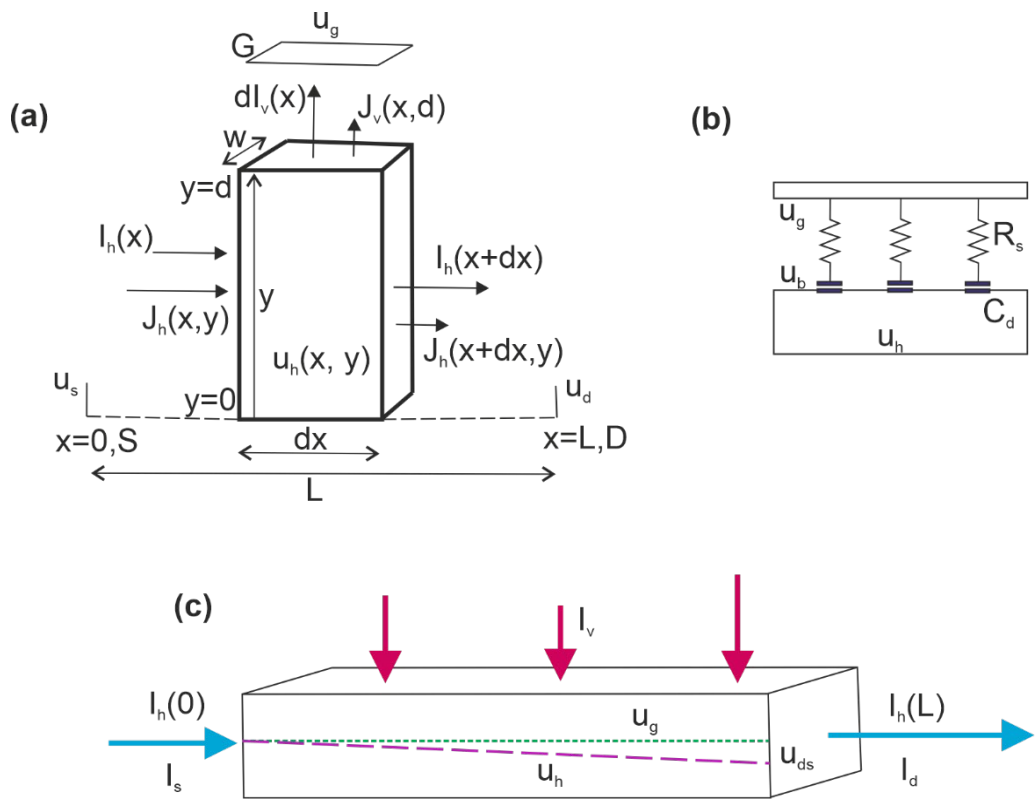

Fig. 1. (a) Scheme of the model for the transport inside the channel and ion exchange with the electrolyte. (b) Additional effect of series resistance in the electrolyte and surface capacitance. (c) By the assumption of charge neutrality, in a transient situation, the difference of horizontal (blue) currents at D and S electrodes, $I_h(L) - I_h(0)$ equals the total ion current entering the channel film (red currents). The voltage distribution in a specific condition measurement in quasiequilibrium conditions with $u_{ds} < 0$ is indicated.

The model shown in Fig. 1 considers a horizontal channel in direction $x$ that spans from 0 to $x = L$, with horizontal current $I_h$ and local voltage $u_h$. The vertical direction in the channel is $y$. Note that current exiting the drain is considered positive. If $z$ is the effective cation density, then by local electroneutrality, the local hole density is

$$p = p_0 - z \tag{1}$$

where $p_0$ in an intrinsic density due to doping. The density $z$ depends on the concentrations of intercalated cations $m$ and anions $a$

$$z = m - a \tag{2}$$

In a volume element of length $dx$ the interface with the electrolyte is at $y = d$. We define an average concentration per unit horizontal distance by the vertical integration of Eq. (1) and (2)

$$Z = w \int_0^d z\, dy = w \int_0^d m\, dy - w \int_0^d a\, dy \tag{3}$$

Hence

$$Z = M - A \tag{4}$$

and

$$P = w \int_0^d p\, dy = P_0 - Z \tag{5}$$

An equilibrium concentration of ions

$$Z_{eq} = Z(u_g) \tag{6}$$

is obtained when the film is charged homogeneously at the gate potential $u_g$. This function can be obtained by electrochemical methods for a given type of film[38-42] as commented in Sec. 2.4.

Two different OECTs operation modes are called depletion and accumulation.[27] In accumulation, the device is normally OFF in the absence of a gate bias (i.e., the semiconducting polymer is initially in its neutral stage). In depletion the active material in the channel is initially doped and the OECT is naturally in its ON state and is turned OFF upon application of a gate voltage (dedoping of the channel). Below we consider two types of situations:

(i) In accumulation (undoped semiconductor) $P_0 = 0$, $Z = -A$, $P = A$.
(ii) In depletion (doped semiconductor) $A = 0$, $Z = M$, $P = P_0 - M$.

### 2.2. Definition of currents

The horizontal flux of hole carriers is given by the Ohm law

$$J_h(x,y) = -p(x,y)\mu_p \frac{du_h}{dx} \tag{7}$$

Here $\mu_p$ is the mobility. The horizontal current is

$$I_h(x) = q\, w \int_0^d J_h(x,y)\, dy \tag{8}$$

where $q$ is the elementary charge. Therefore

$$I_h(x) = -qP(x)\mu_p \frac{du_h}{dx} \tag{9}$$

In Eq. (9) the model is restricted to situations in which the vertical gradient of $u_h$ is not large.

The current across the top interface of the volume element is exclusively ionic. An outward flux of positive ions $J_v^m$ and negative ions $J_v^a$ produces a vertical current $dI_v$ due to the area $w\, dx$, of value

$$dI_v(x) = q\, w\bigl(J_v^m(x,d) - J_v^a(x,d)\bigr)\, dx \tag{10}$$

The current conservation in the volume element of Fig. 1a is

$$I_h(x+dx) - I_h(x) + dI_v(x) = 0 \tag{11}$$

Therefore

$$\frac{\partial I_h}{\partial x} = -qw \int_0^L \bigl(J_v^m(x,d) - J_v^a(x,d)\bigr) dx \tag{12}$$

The simple application of Kirchhoff rules produces the two classical equations of a transmission line model, (10) and (12).[43] This approach has been developed in field-effect transistors[23] and in OECT.[24-27]

The model requires a completion by stating the vertical flux, $J_d$. The flux depends on the voltage in the electrolyte just outside the interface, that is the gate voltage $u_g$, and also on the voltage inside the film, $u_h$. This is developed in Section 2.5.

### 2.3. Capacitive coupling

In the previous theories of OECTs the ion concentration inside the film $z$ is coupled capacitively to the voltage difference across the interface. The total ionic charge is $q\, w\, d\, dx\, z$. Considering the double layer capacitance per unit area in the top surface, $c_d$, we have[26,29,31]

$$q\, d\, z(x)\, w\, dx = c_d\bigl(u_g - u_h(x)\bigr)\, w\, dx \tag{13}$$

In steady state, the horizontal current is constant, $I_h(x) = I_{ds}$, and the integration of Eq. (9) with $u_{ds} = u_h(x=L) - u_h(x=0)$ gives

$$I_{ds} = -\frac{q\, d\, w\, \mu}{L}\left[p_0 - \frac{c_d}{qd}\left(u_g - \frac{1}{2}u_{ds}\right)\right]u_{ds} \tag{14}$$

This is a standard result.[29,30] Later, it was discovered[19,30,44] that in OECT the capacitance $c_d$ shows a correlation with length. Then the volume capacitance

$$c^* = \frac{c_d}{d} \tag{15}$$

is the constant quantity, due to the fact that the ions fill the whole element in Fig. 1. This observation leads one to consider the related ideas of ion intercalation in battery and electrochromic materials[45-47] and in conducting polymers,[39,40] that have been broadly investigated. But then, a different formalism than Eq. (13) is required to account for the intercalation of ions from the electrolyte, based on the ion diffusion concept.[48] Diffusion itself generates a chemical capacitance due to the intercalation process,[49-51] so that the capacitive effect arises naturally.

### 2.4. The chemical capacitance and the thermodynamic equilibrium functions

As a preliminary step to the generalization of the channel dynamics including ionic diffusion, it is important to remark the idea of the chemical capacitance, that appears in the diffusion transport.[52] The chemical capacitance of the ions[49,50] is obtained by the derivative of the thermodynamical function of the ion density with respect to the electrochemical potential.

$$c_\mu = q \frac{dZ_{eq}}{du_g} \tag{16}$$

This equation expresses the volume capacitance $c^*$ mentioned in (15) in the more general denomination of the chemical capacitance $c_\mu$ that is measured in the electrochemistry of organic conductors.[53-55] The bulk origin of the ionic capacitance is well established in organic films.[56] Furthermore it is obtained a relation of the chemical capacitance and the density of states (DOS) $g$,[54,55] as follows

$$c_\mu = q\, g(qu_g) \tag{17}$$

Therefore, the chemical capacitance gives a direct measure of the DOS. It also provides a figure of merit of OECT.[44]

For an anion density that decreases at increasing $u_g$ the chemical capacitance is

$$c_\mu = -q \frac{dA_{eq}}{du_g} \tag{18}$$

The sign in Eq. (18) is due to the fact that when the potential $u_g$ becomes more negative, the concentration of anions increases, so that $c_\mu$ is positive.

Many forms of $Z_{eq}(u_g)$ are possible according to the properties of the organic film.[41,42,45,46,53,57,58] In coherence with the drift-diffusion simulations described later, here we use the density of states (DOS) for inserted anions

$$g_a(qu_g) = -\frac{dA_{eq}}{d(qu_g)} = \frac{A_0}{k_B T}\left[\frac{q(u_v-u_g)}{k_B T}\right]^{1/2} \tag{19}$$

The $k_B$ is the Boltzmann's constant, $T$ the absolute temperature, $u_v$ the valence edge potential, $A_0$ a density per length, from the volume density $a_0 = A_0/wd$. This is the DOS used for holes in the simulations below, which by electroneutrality it applies to ions as well.

Accordingly the thermodynamic function is

$$A_{eq}(u_g) = -\int_{u_g}^{u_v} g_a(x)dx = \frac{2}{3}A_0\left[\frac{q(u_v-u_g)}{k_B T}\right]^{3/2} \tag{20}$$

For insertion of cations

$$g_m(qu_g) = -\frac{dM_{eq}}{d(qu_g)} = \frac{M_0}{k_BT}\left[\frac{q(u_g-u_c)}{k_BT}\right]^{1/2} \tag{21}$$

$M_0$ is a density per length, from the volume density $m_0 = M_0/wd$. For a cation density that increases at increasing $u_g$ the chemical capacitance

$$c_\mu = q\frac{dM_{eq}}{du_g} \tag{22}$$

is positive.

The thermodynamic function is

$$M_{eq}(u_g) = -\int_{u_c}^{u_g} g_m(x)dx = \frac{2}{3}M_0\left[\frac{q(u_g-u_c)}{k_BT}\right]^{3/2} \tag{23}$$

The general conclusions on time transient and hysteresis derived below are not dependent on the specific expressions $A_{eq}, M_{eq}$, provided that these functions are such that the chemical capacitances remain positive.

### 2.5. Diffusion of intercalated ions

Here we aim introduce the diffusion dynamics that departs from the previous treatments based on Eq. (13), and has application for solid state or electrochemical ionic diffusion in organic and inorganic materials.[11,12]

The diffusion of ions in the film is described by the conservation equation

$$\frac{\partial z}{\partial t} = -\frac{\partial J_v}{\partial y} \tag{24}$$

and the Fick's law that states that the ion flux is proportional to the ion diffusion coefficient $D_{ion}$ and the gradient of the concentration

$$J_v = -D_{ion}\frac{\partial z}{\partial y} \tag{25}$$

Considering that the bottom layer in Fig. 1a is blocking to ions, $J_v(y=0) = 0$, the integral of (24) gives

$$w\int_0^d \frac{\partial z}{\partial t}dy = -wJ_v(x,d) \tag{26}$$

$$wJ_v(x) = -\frac{\partial Z}{\partial t} \tag{27}$$

This last equation incorporates the assumption that the ionic charge vertically injected is compensated by electronic charge of the opposite sign from the immediate surrounding. This is the universal assumption used in intercalation systems such as Li ion batteries, that are described by the ion diffusion rate at the boundary with the electrolyte.[59] However, this assumption requires that the dielectric relaxation time is short, which occurs when sufficient electronic conductivity is available. In more detailed treatments one should consider the coupling of ionic and electronic local currents.[60,61]

The conservation equation for cations and anions give the results

$$wJ_v^m(x) = -\frac{\partial M}{\partial t} \tag{28}$$

$$wJ_v^a(x) = -\frac{\partial A}{\partial t} \tag{29}$$

From Eqs. (9, 12) the equations of the transmission line model can be stated

$$I_h(x) = -q(P_0 - M + A)\mu_p \frac{du_h}{dx} \tag{30}$$

$$\frac{\partial I_h}{\partial x} = q\left(\frac{\partial M}{\partial t} - \frac{\partial A}{\partial t}\right) \tag{31}$$

### 2.6. Current and electronic transit time

The drain-source voltage is

$$u_{ds} = u_h(L) - u_h(0) \tag{32}$$

For a small $u_{ds}$ that produces a uniform electrical field, the current (30) can be written

$$I_h(x) = -q\mu_p \frac{u_{ds}}{L}(P_0 - M(x) + A(x)) \tag{33}$$

By integration of Eq. (31) we obtain

$$I_h(L) - I_h(0) = q \int_0^L \left(\frac{\partial M}{\partial t} - \frac{\partial A}{\partial t}\right) dx \tag{34}$$

We define drain and source current as

$$I_s = I_h(0) \tag{35}$$

$$I_d = I_h(L) \tag{36}$$

Drain and source currents are not equal in the transient condition, due to the charging vertical current of intercalation, as shown in Fig. 1c.

To confirm the interpretation of signs of Eq. (34) consider an electronically blocking source contact, $I_h(0) = 0$. We inject a pulse of anions and Eq. (34) gives

$$I_h(L) = -q \int_0^L \left(\frac{\partial A}{\partial t}\right) dx \quad \text{(blocking source)} \tag{37}$$

If we increase the concentration of anions, by more negative $u_g$, then $\partial A/\partial t > 0$, hence the current (37) is negative, as holes enter the film from the right side in Fig. 1c.

Let us define the sign of the voltage drop as

$$\theta = \frac{u_{ds}}{|u_{ds}|} \tag{38}$$

The standard operation of p-type transistor corresponds to $u_{ds} < 0$ or $\theta = -1$. In the notation used here, the stationary $I_d$ is positive. Studying both signs of $\theta$ is that, within the scope of this model, it is equivalent to studying both the drain ($\theta = -1$) and source ($\theta = +1$) currents, which is most important for transient studies.[23,62]

We introduce the electronic transit time along the channel length[29]

$$\tau_e = \frac{L^2}{\mu_p |u_{ds}|} \tag{39}$$

The current (33) is expressed

$$I_h(x) = -\theta \frac{qL}{\tau_e}(P_0 - M(x) + A(x)) \tag{40}$$

### 2.7. A simple solution: Bernard-Malliaras model for the undoped situation

Assume an undoped sample in which the hole density is due to intercalated anions $A$. To obtain a functional model for the transient currents, a solution of Eq. (31) and (40) must be obtained. We can combine these equations as

$$\frac{\partial A}{\partial x} = \theta \frac{\tau_e}{L} \frac{\partial A}{\partial t} \tag{41}$$

and we obtain the integral

$$A(L) - A(0) = -\theta \frac{\tau_e}{L} \int_0^L \frac{\partial A}{\partial t} dx \tag{42}$$

Following Bernard and Malliaras[29] we allow for holes to be injected from both contacts and thus we distribute the transient ionic current in two different parts by the fractional constant $0 < f < 1$

$$A(L) = A_{av} - f \frac{\tau_e}{L} \int_0^L \frac{\partial A}{\partial t} dx = A_{av} + \theta f \tau_e \frac{\partial A_{av}}{\partial t} \tag{43}$$

$$A(0) = A_{av} - (1-f) \frac{\tau_e}{L} \int_0^L \frac{\partial A}{\partial t} dx = A_{av} + \theta(1-f)\tau_e \frac{\partial A_{av}}{\partial t} \tag{44}$$

The $A_{av}$ is an average concentration taken by a suitable procedure. Different integration schemes[24,63,64] to justify the value of $f$ are summarized in Ref. [19]. Removing the subscript $av$, the drain current in (43) is

$$I_h(L) = \theta \frac{qL}{\tau_e} A - qfL \frac{dA}{dt} \tag{45}$$

### 2.8. Doped semiconductor

We take the case $A = 0, Z = M, P = P_0 - M$. By similar calculations as before

$$I_h = -\theta \frac{qL}{\tau_e}(P_0 - M) \tag{46}$$

$$\frac{\partial M}{\partial x} = \theta \frac{\tau_e}{L} \frac{\partial M}{\partial t} \tag{47}$$

$$I_h(L) = -\theta \frac{qL}{\tau_e}(P_0 - M) + qfL \frac{dM}{dt} \tag{48}$$

If a pulse $\partial M/\partial t > 0$ is injected a positive hole current $I_d = qfL\partial M/\partial t > 0$ has to leave the channel.

### 2.9. Local diffusion effect

The previous models proposed in the literature, leading to Eq. (45), have not considered explicitly the ion diffusion process. For a rigorous solution, one can solve Eq. (12) coupled to Eq. (25). This is a complex problem that will generate another (vertical) transmission line at each point $x$.[65]

To obtain a physically meaningful but approximated solution, we have suggested[32] that the boundary flux is given by a gradient of concentration associated to different electrochemical potentials of the volume element of Fig. 1a, from the bottom $u_h$ to the top $u_g$. Then

$$J_v^a(x,d) = -D_{ion}\frac{\partial a}{\partial y} = \frac{D_X}{d}[a(u_h) - a(u_g)] \tag{49}$$

Therefore

$$\frac{\partial A}{\partial t} = \frac{D_{ion}}{d^2}[A(u_g) - A(u_h)] \tag{50}$$

We also define the transit time for diffusion

$$\tau_d = \frac{d^2}{D_{ion}} \tag{51}$$

Using the notion that the chemical capacitance is charged by a diffusion current, we can express Eq. (29) as

$$\frac{\partial A}{\partial t} = \frac{1}{\tau_d}(A_{eq}(u_g) - A(u_h)) \tag{52}$$

Eqs. (45) and (52) are the model introduced in Ref. [32], that here has been rigorously justified.

Similarly for the insertion of cations

$$\frac{\partial M}{\partial t} = \frac{1}{\tau_d}(M_{eq}(u_g) - M(u_h)) \tag{53}$$

Note that the diffusion of ions in Eq. (49) requires a flow of compensating holes. If the time associated with hole compensation is $\tau_v$ then it is required that $\tau_v \ll \tau_d$, as commented earlier, to satisfy the electroneutrality condition. If only a vertical flow is needed for the charge compensation then

$$\tau_v = \frac{d^2}{\mu_p(u_g - u_h)} \tag{54}$$

Based on the general transmission line approach of Eqs. (9, 12) one can add different effects that influence the transient behaviour, for example a distributed interfacial capacitance as Eq. (13)[24,66] and a series resistance in the electrolyte.[25] The realistic 2D flows of charge will be discussed in the simulations in Section 6.

## 3. Transient switching dynamics

We start the study of transient behaviour based on the model of Eqs. (45, 52).[32] Let us summarize the structure of the model. We consider an accumulation semiconductor, $P_0 = M = 0$, for a $u_{ds}$ small voltage, in which $A(x,t)$ is nearly homogeneous, with equilibrium function $A_{eq}$ given in Eq. (20). In the stationary dc current the cation density obtains the value of equilibrium according to the gate voltage $A_{eq}(u_g)$. As mentioned before the voltage applied to the gate is $u_g$. The voltage in the channel is $u_h$. The model for any time-dependent situation consists on two equations for the three variables $I_d, u_g, A$.

$$I_d = -\theta\frac{qL}{\tau_e}A - qLf\frac{dA}{dt} \tag{55}$$

$$\tau_d\frac{dA}{dt} = A_{eq}(u_g) - A \tag{56}$$

### 3.1. The linear equations for a small perturbation

The properties of switching will be first investigated by a small step of the external voltage $\Delta V$ at $t = 0$ that changes the gate voltage from an initial value $u_{g0}$

$$u_g = u_{g0} + \Delta V \tag{57}$$

The internal voltage in the film changes as

$$u_h(t) = u_{g0} + \Delta v(t) \tag{58}$$

where $\Delta v(t = 0) = 0$ and $\Delta v(t = \infty) = \Delta V$. Considering the chemical capacitance of Eq. (18), the change of the ion densities in Eqs. (55-56) is

$$A(u_{g0} + \Delta v, t) = A_{eq}(u_{g0}) - \frac{c_\mu}{q}\Delta v(t) \tag{59}$$

$$A_{eq}(u_{g0} + \Delta V) = A_{eq}(u_{g0}) - \frac{c_\mu}{q}\Delta V \tag{60}$$

Thus Eq. (56) takes the form

$$\tau_d \frac{d\Delta v}{dt} = \Delta V - \Delta v \tag{61}$$

The total step of the current is

$$I_{fin} = I_{in} + \theta \frac{c_\mu L}{\tau_e} \Delta V \tag{62}$$

with $I_{fin}$ and $I_{in}$ being the final and initial currents, respectively, where

$$I_{in} = -\theta \frac{qL}{\tau_e} A_{eq}(u_{g0}) \tag{63}$$

A total quantity of electronic charge

$$\Delta Q = \theta L c_\mu \Delta V \tag{64}$$

enters the channel for the new equilibrium situation. Now the time-dependent current of Eq. (55) can be expressed

$$I_d = -\theta \frac{qL}{\tau_e} A_{eq}(u_{g0}) + \theta L c_\mu \frac{1}{\tau_e} \Delta v + L c_\mu f \frac{d(\Delta v)}{dt} \tag{65}$$

And using Eq. (61) we have

$$I_d = -\theta \frac{qL}{\tau_e} A_{eq}(u_{g0}) + \theta L c_\mu \frac{1}{\tau_e} \Delta v + L c_\mu f \frac{1}{\tau_d}(\Delta V - \Delta v) \tag{66}$$

The transient current is composed of two parts. The second term corresponds to the horizontal transport of the injected charge. The third term is the compensation of the vertical injection of the charge. Both terms, when combined, inject the total charge (64).

Let us write the Eq. (66) in the form

$$I_d = -\theta \frac{qL}{\tau_e} A_{eq}(u_{g0}) + L c_\mu f \frac{1}{\tau_d} \Delta V + L c_\mu \left(\frac{\theta}{\tau_e} - \frac{f}{\tau_d}\right) \Delta v \tag{67}$$

We remark that the current step at the first instant is

$$\Delta I_d(t = 0) = +L c_\mu f \frac{1}{\tau_d} \Delta V \tag{68}$$

We can analyze these effects by using the explicit dependence on time. Eq. (61) gives

$$\Delta v = \Delta V \left(1 - e^{-t/\tau_d}\right) \tag{69}$$

Therefore

$$I_d(t) = -\theta \frac{qL}{\tau_e} A_{eq}(u_{g0}) + Lc_\mu f \frac{1}{\tau_d} \Delta V + Lc_\mu \left(\frac{\theta}{\tau_e} - \frac{f}{\tau_d}\right) \Delta V \left(1 - e^{-t/\tau_d}\right) \qquad (70)$$

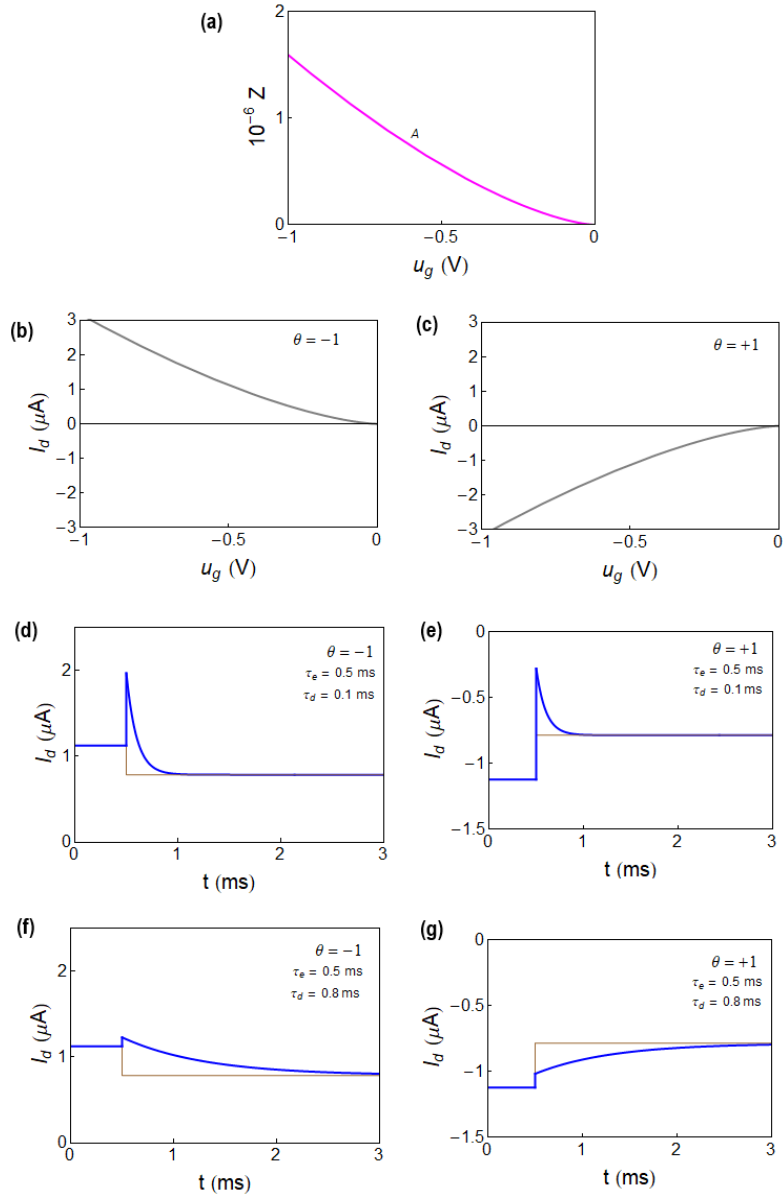

Fig. 2. Undoped semiconductor. (a) Equilibrium anion density in the film. The stationary current (b, c) according to the sign $\theta = u_{ds}/|u_{ds}|$ and (d-g) the possible four different types of transient response with respect to time, for a step voltage at $V_0 = -0.5\ V$ and $\Delta V = 0.1\ V$ at $t_0 = 0.5$ ms. The brown line is an immediate current response. We take parameters of Table 1, $\tau_e = 0.5$ ms, $a_0 = 6.2 \times 10^{18}$ cm$^{-3}$, $A_0 =$

$6.2 \times 10^{12}$ m⁻¹, $u_v = 0$, $f = 0.5$ and $\tau_d$ as indicated.

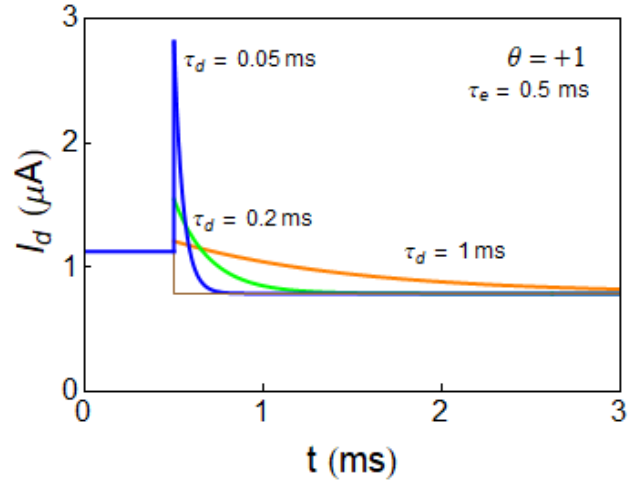

Fig. 3. Switching transients for an undoped semiconductor, same as in Fig. 2 for several values of $\tau_d$.

Table 1

| | | |
|---|---|---|
| channel length | $L$ | 50 μm |
| thickness | $d$ | 100 nm |
| width | $w$ | 10 μm |
| Hole mobility | $\mu_p$ | 0.02 cm²/Vs |
| Source-drain voltage | $|u_{ds}|$ | 0.1 V |
| Thermal energy | $k_B T$ | 0.026 V |

In Fig. 2 we show the transient response for the parameters outlined in Table 1. There are four different cases, distinguished by $\tau_e < \tau_d$ or vice versa, and also by the sign of the current $\theta$, that makes the step either increase or decrease the current in the final state. The initial spike is only dependent on $\Delta V / \tau_d$, by Eq. (68). The characteristic time for the transient is the vertical ion diffusion time $\tau_d$. This is shown in several cases in Fig. 3 The area of the curve corresponds to (64), so that the in Fig. 3 spike is higher for shorter charging time $\tau_d$. The are two possibilities in the transient term $\Delta v$ of Eq. (67) and the system will choose the fastest. If $\tau_e < \tau_d$, Fig. 2g, the initial spike is smaller than the equilibrium current, hence the current does not make a spike but increases since the first instant.[29]

### 3.2. Doped semiconductor

Here $A = 0$, $Z = M$, $P = P_0 - M$. The equations for the general transient behaviour are

$$I_h(L) = -\theta \frac{qL}{\tau_e}(P_0 - M) + qfL\frac{dM}{dt} \tag{71}$$

$$\tau_d \frac{dM}{dt} = M_{eq}(u_g) - M \tag{72}$$

For a small voltage step $\Delta V$ the internal voltage in the film changes as

$$u_h(t) = u_{g0} + \Delta v(t) \tag{73}$$

The total step of the current is

$$I_{fin} = I_{in} + \theta \frac{c_\mu L}{\tau_e} \Delta V \tag{74}$$

Here $I_{in}$ is the initial current,

$$I_{in} = -\theta \frac{qL}{\tau_e}(P_0 - M_{eq}) \tag{75}$$

The time-dependent current is

$$I_d = -\theta \frac{qL}{\tau_e}\left(P_0 - M(u_{g0})\right) + \theta L c_\mu \frac{1}{\tau_e}\Delta v + f L c_\mu \frac{d(\Delta v)}{dt} \tag{76}$$

This can be written

$$I_d = -\theta \frac{qL}{\tau_e}\left(P_0 - M(u_{g0})\right) + f L c_\mu \frac{1}{\tau_d}\Delta V + L c_\mu \left(\frac{\theta}{\tau_e} - \frac{f}{\tau_d}\right)\Delta v \tag{77}$$

Therefore

$$I_d(t) = -\theta \frac{qL}{\tau_e}\left(P_0 - M(u_{g0})\right) + f L c_\mu \frac{1}{\tau_d}\Delta V + L c_\mu \left(\frac{\theta}{\tau_e} - \frac{f}{\tau_d}\right)\Delta V(1 - e^{-t/\tau_d}) \tag{78}$$

The initial step is again given by Eq. (68)

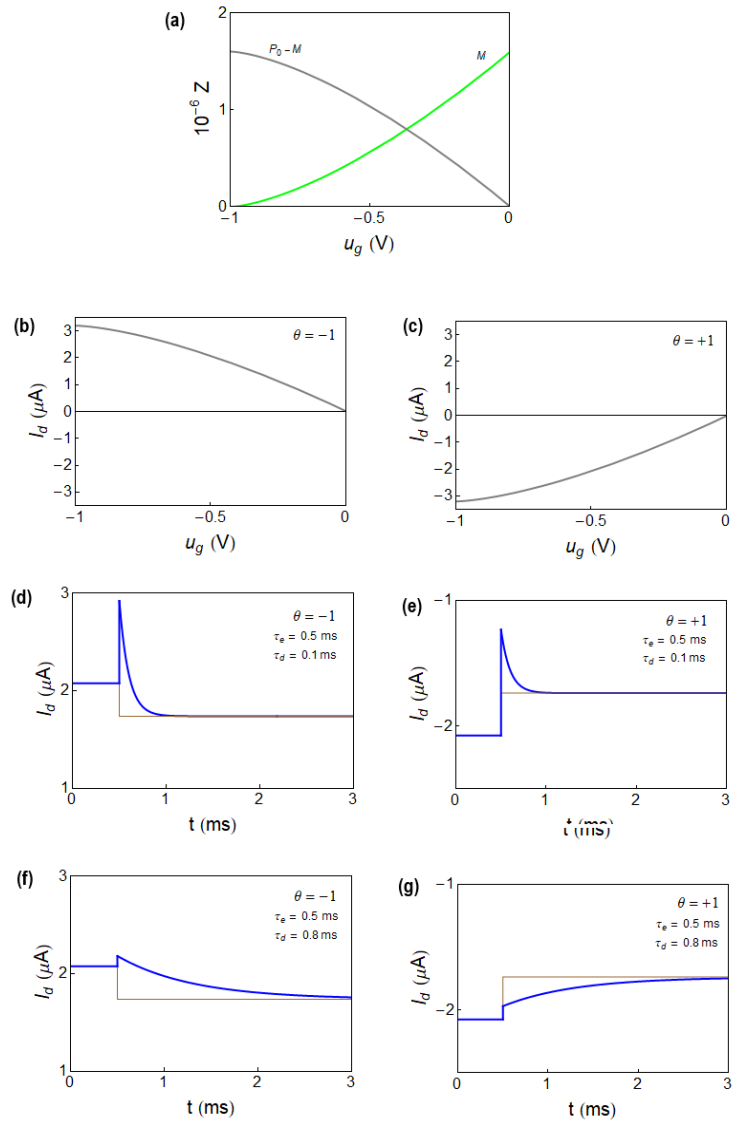

Fig. 4. Doped semiconductor. (a) Ionic density. (b, c) The stationary current according to the sign $\theta = u_{ds}/|u_{ds}|$ and (d-h) the possible four different types of transient response with respect to time, for a step voltage at $V_0 = -0.5\ V$ and $\Delta V = 0.1\ V$. The brown line is an immediate current response. We take parameters of Table 1, $u_c = -1$ V, $\tau_e = 0.5$ ms, $m_0 = 6.2 \times 10^{18}$ cm$^{-3}$, $M_0 = 6.2 \times 10^{12}$ m$^{-1}$, $P_0 = 160\ M_0$, $f = 0.5$ and $\tau_d$ as indicated.

The four possible types of decays are shown in Fig. 4. For better comparison of doped and undoped cases we set the DOS parameter $u_c = -1$ V so that $I_d = 0$ at the origin. The depletion case is obtained by shifting the curves to the right by $u_c = 0$ V.

### 3.3. General transient dynamics and series resistance

The transient decays can be investigated for a large perturbation, based on the Eqs. (55-56) or (69-70).[64,67] We can add the effect of a series resistance $R_s$ and the surface capacitance of Eq. (13) as indicated in Fig. 1b, where $u_b$ is the potential at the boundary of the channel and the electrolyte. The transient equations become

$$C_d \frac{du_b}{dt} = \frac{1}{R_s}(u_g - u_b) \quad (79)$$

$$I_d = -\theta \frac{qL}{\tau_e} A(u_h) - qLf \frac{dA}{dt}(u_h) \quad (80)$$

$$\tau_d \frac{dA}{dt}(u_h) = A_{eq}(u_b) - A(u_h) \quad (81)$$

The results are shown in Fig. 5, and are in agreement with the transients reported.[64]

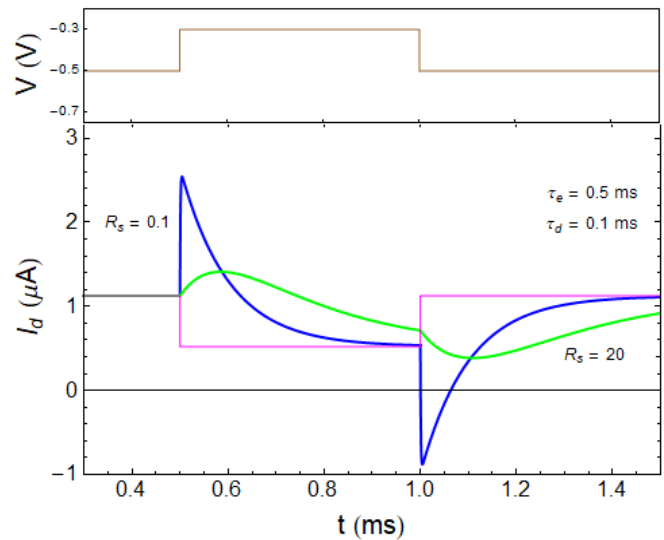

Fig. 5. Undoped semiconductor. Transient response for a step voltage at $V_0 = -0.5\ V$ and $\Delta V = 0.1\ V$ and different values of the series resistance (in ohm). The brown line is the applied voltage, the purple line is the correspondent equilibrium current. We take parameters of Fig. 2 and $C_d = 0.01$ F.

## 4. Hysteresis effects
### 4.1. Undoped semiconductor

Hysteresis in current-voltage transfer curves is usually analysed applying the time dependent voltage with constant velocity $v_r$,

$$u_g = v_r t \tag{82}$$

for a go and return cycle in the voltage range of interest. However, in practice the instrument develops small steps of voltage at time steps $\Delta t = \Delta V / v_r$, as shown in Fig. 6 for $\theta = -1$ and in Fig. 7 for $\theta = +1$.

From the transient response in Fig. 2 we can deduce the hysteresis effect that will happen due to ionic charging. As we have mentioned the time of equilibration is $\tau_d$ in all cases. If the time step is long, the current reaches equilibrium at each time, and there is no hysteresis as in Fig. 6a. Now if we apply a voltage step, but at time $t \approx \tau_d$ apply a second step, the current in Fig. 2c obtains an excess with respect to the equilibrium line. This is accumulated up to the red arrow as shown in Fig. 6c. Hence in the forward direction the measured current will be *larger* than equilibrium current, and the hysteresis current will be larger when *higher* sweep velocity is applied, forming capacitive hysteresis.[68]

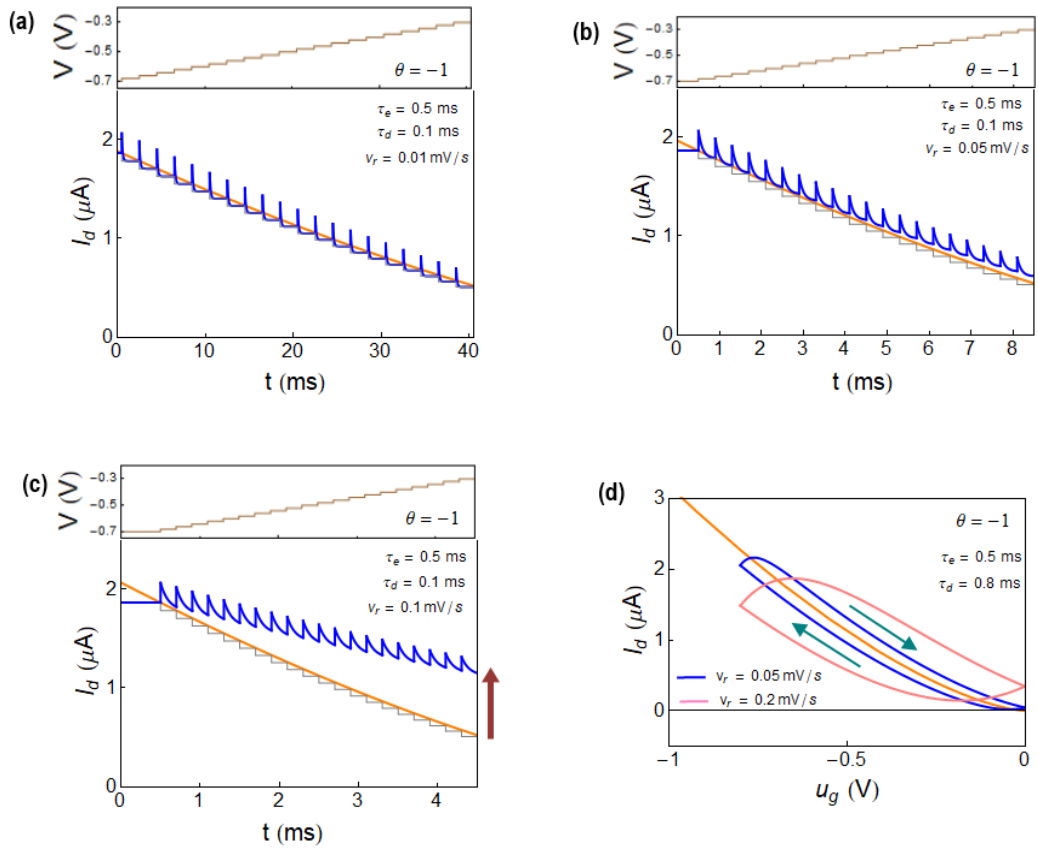

Fig. 6. Undoped semiconductor for $\theta = -1$ ($u_{ds}<0$). (a-c) Increasing the drain current by small voltage steps $\Delta V = 0.02\ V$ starting at $V_0 = -0.7$ V ($t_0 = 0.5$ s) with velocity $v_r$ at time steps $\Delta t = \Delta V/v_r$, at increasingly faster rates from (a) to (c). The grey step line is an immediate current response, the orange line is the equilibrium current-voltage curve, and the red arrow in (c) is the final hysteresis current. (d) Continuous hysteresis loops at the indicated scan rates. Same parameters as Fig. 2 and $\tau_d, \tau_e$ as indicated.

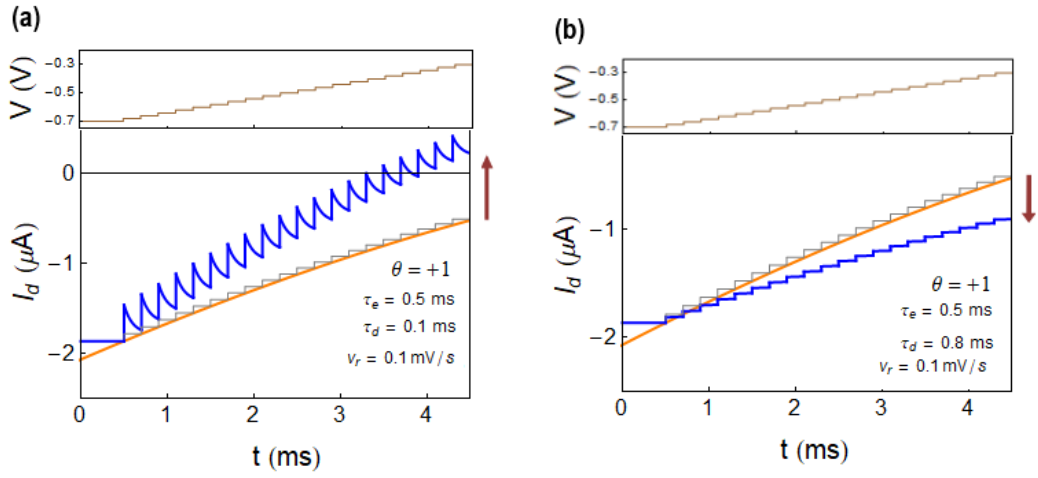

Fig. 7. Undoped semiconductor for $\theta = +1$. (a-c) Increasing the drain current by small voltage steps $\Delta V = 0.02\ V$ starting at $V_0 = -0.7$ V ($t_0 = 0.5$ s) with velocity $v_r$ at time steps $\Delta t = \Delta V/v_r$, at increasingly faster rates from (a) to (b). The grey step line is an immediate current response, the orange line is the equilibrium current-voltage curve, and the red arrow in (c) is the final hysteresis current. Same parameters as Fig. 2 and $\tau_d, \tau_e$ as indicated.

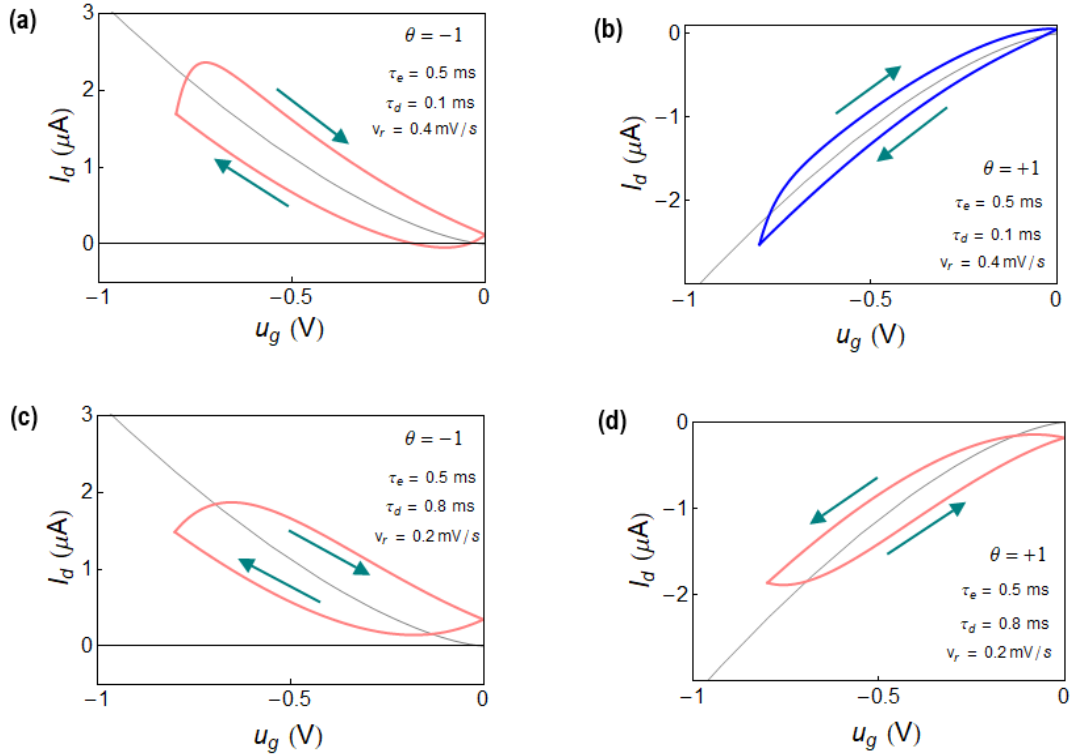

Fig. 8. Undoped semiconductor. Continuous hysteresis loops at the indicated scan rates. Same parameters as Fig. 2 and $\theta, \tau_d, \tau_e$ as indicated.

Let us formulate the model equations (55, 56) under the perturbation (82), as follows

$$I_d = -\theta \frac{qL}{\tau_e} A(u_g) - qLf \frac{dA}{dt} \tag{83}$$

$$\tau_d v_r \frac{dA}{du_g} = A_{eq}(u_g) - A(u_g) \tag{84}$$

The $I - u_g$ curves are obtained by integration of Eq. (84) to find $A(u_g)$, then plotting the current in Eq. (83). The hysteresis loops corresponding to Fig. 6b, c, are shown in Fig. 6d. The hysteresis effect becomes larger at increasing scan rate.

The four types of loops according to the four decay types of Fig. 2 (depending on the sign of the $u_{ds}$ and the dominant relaxation time) are shown in Fig. 8. We note that in Fig. 8a, b, c, (marked in pink line) the hysteresis is similar as it consists on an increase of the absolute value of the current in the forward direction, with respect to the equilibrium current, and a decrease in the reverse direction. This is the case of capacitive hysteresis in the general classification.[32,68] However Fig. 7b and 8b show a different behaviour (marked in blue line), in which the current at forward bias is less than the equilibrium current. This is the case of inductive hysteresis.[32,68]

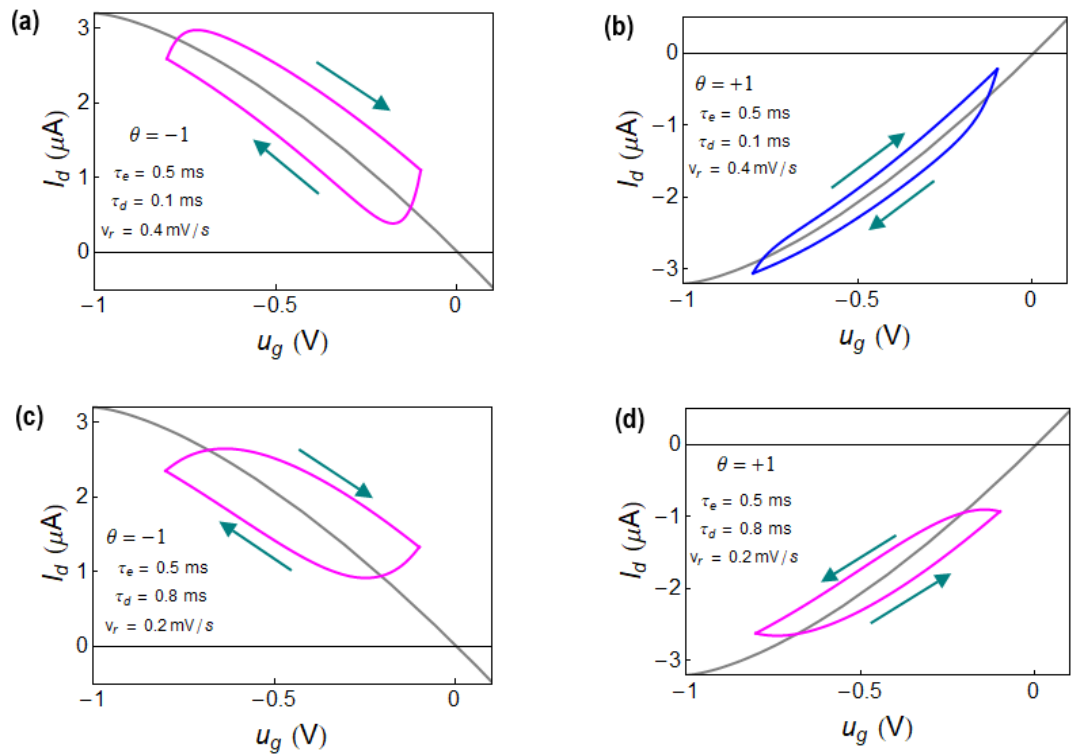

Fig. 9. Doped semiconductor. Continuous hysteresis loops at the indicated scan rates. Same parameters as Fig. 4 and $\theta, \tau_d, \tau_e$ as indicated.

For the doped semiconductor the hysteresis curves are obtained solving the equations

$$I_h(L) = -\theta \frac{qL}{\tau_e}(P_0 - M(u_g)) + qfL\frac{dM}{dt} \qquad (85)$$

$$\tau_d v_r \frac{dM}{du_g} = M_{eq}(u_g) - M(u_g) \qquad (86)$$

The results corresponding to the step decays of Fig. 4 are shown in Fig. 9. In correspondence with the undoped case, we obtain three cases of capacitive hysteresis, a, c, d (purple), and one case of inductive hysteresis, b, (blue).

## 5. Experimental results

According to the electrochemical model equations (83) and (84), four types of hysteresis loops in $I_d$ vs $u_g$ plots (transfer curves) were classified, which are determined by the sign of the voltage drop $u_{ds}$ ($\theta$) and the ratio of characteristic times $\tau_e$ and $\tau_d$. In order to experimentally validate the feasibility of hysteresis loops in real devices, the transfer curves and current transients of OECTs based on undoped semiconductor under various conditions were measured. For this purpose, OECTs with the channel based on a P3HT layer were fabricated and characterized.

P3HT, a well-studied thiophene-based conjugated polymer, has found applications in various electronic devices, including solar cells[69,70], LEDs[71], and organic transistors[72,73]. To construct the channel of an Organic Electrochemical Transistor (OECT), a layer of P3HT was cast onto a glass substrate with pre-patterned ITO drain/source electrodes using a spin-coating technique. The spin-coated P3HT layer was intentionally not annealed to maintain its amorphous structure. According to research by Ginger and colleagues[72,73], a transistor based on amorphous P3HT functions as an electrochemical transistor, whereas a highly crystalline P3HT bulk can impede ion intake, leading to the device operating as an electrolyte-gated field-effect transistor.

### 5.1. Methods

To fabricate organic electrochemical transistors (OECTs) operating in accumulation mode, Poly(3-hexylthiophene) (P3HT) was utilized as the channel material. P3HT (LT-S909) was procured from Luminescence Technology Corp. Interdigitated pre-patterned ITO glass substrates (S161: Width × Length: 30 mm × 50 μm) from Ossila served as the source-drain electrodes.

The P3HT solution was prepared by dissolving 25 mg of polymer in 1 ml of chlorobenzene within a glovebox with an inert atmosphere. The solution was stirred for 3 hours at 45°C before spin-coating. Prior to use, the ITO-patterned source-drain substrates underwent rigorous cleaning. Initially, the substrates were sonicated in deionized (DI) water with detergent for 10 minutes, followed by rinsing three times with DI water. Subsequently, they were sonicated in acetone and isopropyl alcohol (IPA) for 10 minutes each. Finally, the substrates were dried using nitrogen flow and treated with UV-ozone for 15 minutes to remove any residual organics and enhance substrate surface wettability.

P3HT channel layers were deposited using a spin-coating technique. A volume of 30 μL of the P3HT solution at 45°C was dispensed onto the substrate spinning at a rate of 1000 rpm and kept rotating for 1 minute. The as-cast P3HT films were utilized without annealing, maintaining them in an amorphous state. Previous studies have shown that amorphous channel layers facilitate the penetration of anions under appropriate gate bias voltage, while organic transistors based on crystalline P3HT layers primarily function as electrolyte-gated transistors.

A 10 mM or 1mM KCl aqueous solution was employed as the electrolyte, with an Ag wire serving as the gate electrode. IV curves were measured using a 2-channel Keithley source meter controlled by a customized LabVIEW program. We restricted our measurement of transfer curves to a voltage range of $0 \text{ V} > V_{ds} > -0.2 \text{ V}$ to minimize the effect of drain (or source) potential and ensure relatively uniform doping along the channel length.

### 5.2. Results

Transfer curves of P3HT OECTs were obtained by sweeping the gate bias ($u_g$) from -0.5 V to 0.1 V at various scan rates ($v_r$). The transfer curves were probed at different drain biases to adjust the characteristic times $\tau_e$, which are inversely proportional to $V_{ds}$. To ensure that the device remained in the linear regime, transfer curves were measured at low $V_{ds}$ (ranging from 25 mV to 200 mV).

The transfer curves measured at various values and polarities of $V_{ds}$ (at $\theta = -1$ and $\theta = +1$) are depicted in Figure 10. As seen in Figure 10a, for $\theta = -1$, the hysteresis loops exhibit a capacitive trend at all values of $V_{ds}$ ranging from -25 mV to -200 mV, which is consistent with the theoretical prediction in Section 4. As demonstrated by plotting equations (83) and (84) in Figure 8a and 8b, for both ratios of electronic and ionic characteristic times ($\tau_e > \tau_d$ or vice versa), for $\theta = -1$, hysteresis loops exhibit only capacitive behavior.

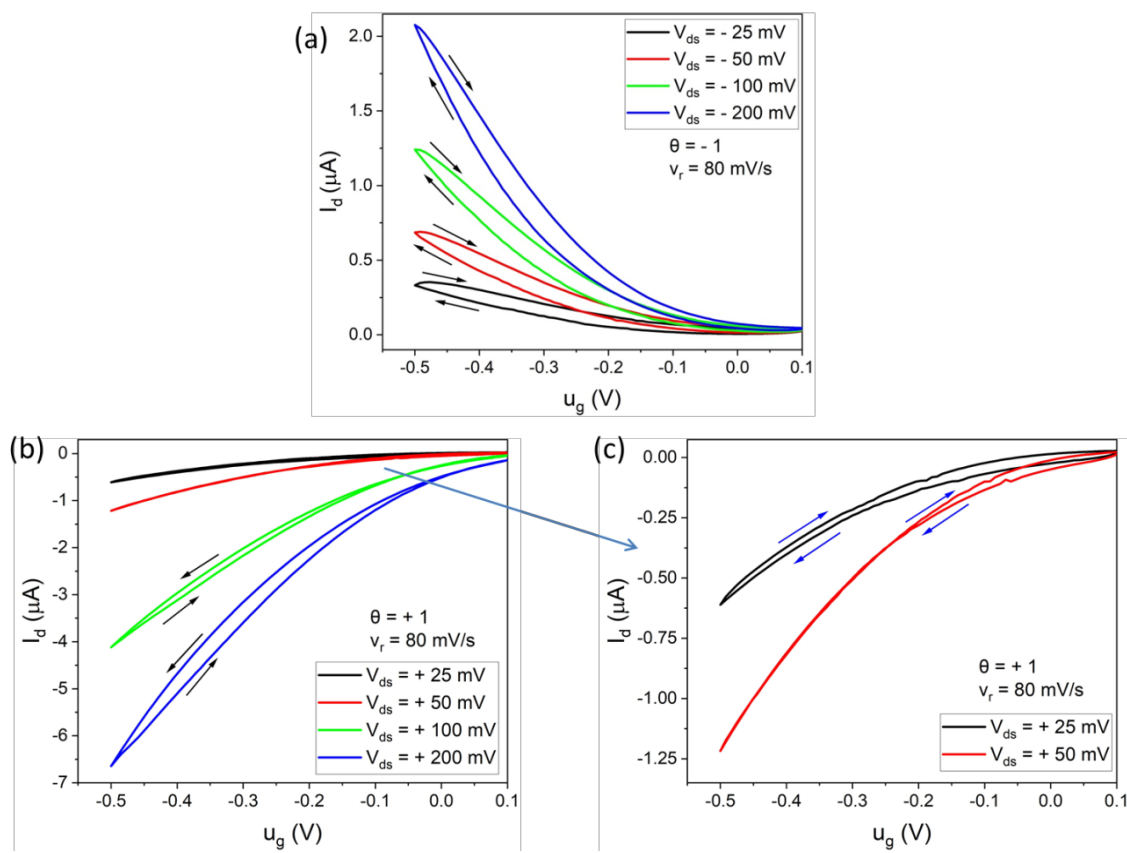

Fig. 10. Transfer curves of P3HT OECT (the accumulation OECT with the initially undoped channel) at different values of drain-source voltage . (a) for $\theta = +1$; (b) for $\theta = -1$; (c) magnified (b) plots.

In the case of $\theta = +1$, the hysteresis model predicts the possibility of observing an inductive hysteresis in the transfer curve of the OECT. According to Figure 8b, inductive hysteresis appears only for $\theta = +1$ when the electronic relaxation time is dominant ($\tau_e > \tau_d$). Experimentally, this condition can be achieved by increasing the channel length and decreasing $V_{ds}$ (Eq. (39)), or by decreasing the channel thickness and increasing an anion diffusion coefficient (Eq. (51)). In our study, all parameters affecting $\tau_e$ and $\tau_d$ were fixed, except $V_{ds}$. The electronic characteristic time is inversely proportional to $V_{ds}$, and by decreasing $V_{ds}$, the condition of $\tau_e > \tau_d$ can be possibly reached.

Indeed, as shown in Figure 10b and 10c, for $\theta = +1$, the decrease of $V_{ds}$ switches the hysteresis type. At relatively high $V_{ds}$, we observe capacitive hysteresis in the transfer curve, whereas at very low $V_{ds}$ (at 25 mV), it becomes inductive in accordance with the analytical model. In fact, we observe a shrinkage of capacitive hysteresis with decreasing $V_{ds}$. However, it should be noted that this phenomenon also can be explained by residual doping of P3HT, which will be discussed in detail in the Sec. 6.

In the case of $\theta = +1$, as shown in Figure 11c, inductive hysteresis was detected at very low $V_{ds}$. However, it also depends on the scan rate ($v_r$). Inductive hysteresis is clearly observed at higher $v_r$. As seen in Figure 12, at relatively low $v_r$ and low $V_{ds}$, the hysteresis exhibits a capacitive pattern. At a fixed $V_{ds}$, the capacitive hysteresis switches to inductive hysteresis at higher $v_r$. When $V_{ds}$ is maintained at 50 mV, the switch between types of hysteresis is observed at $v_r$ = 62 mV/s, and at $v_r$ = 120 mV/s, the hysteresis is predominantly inductive.

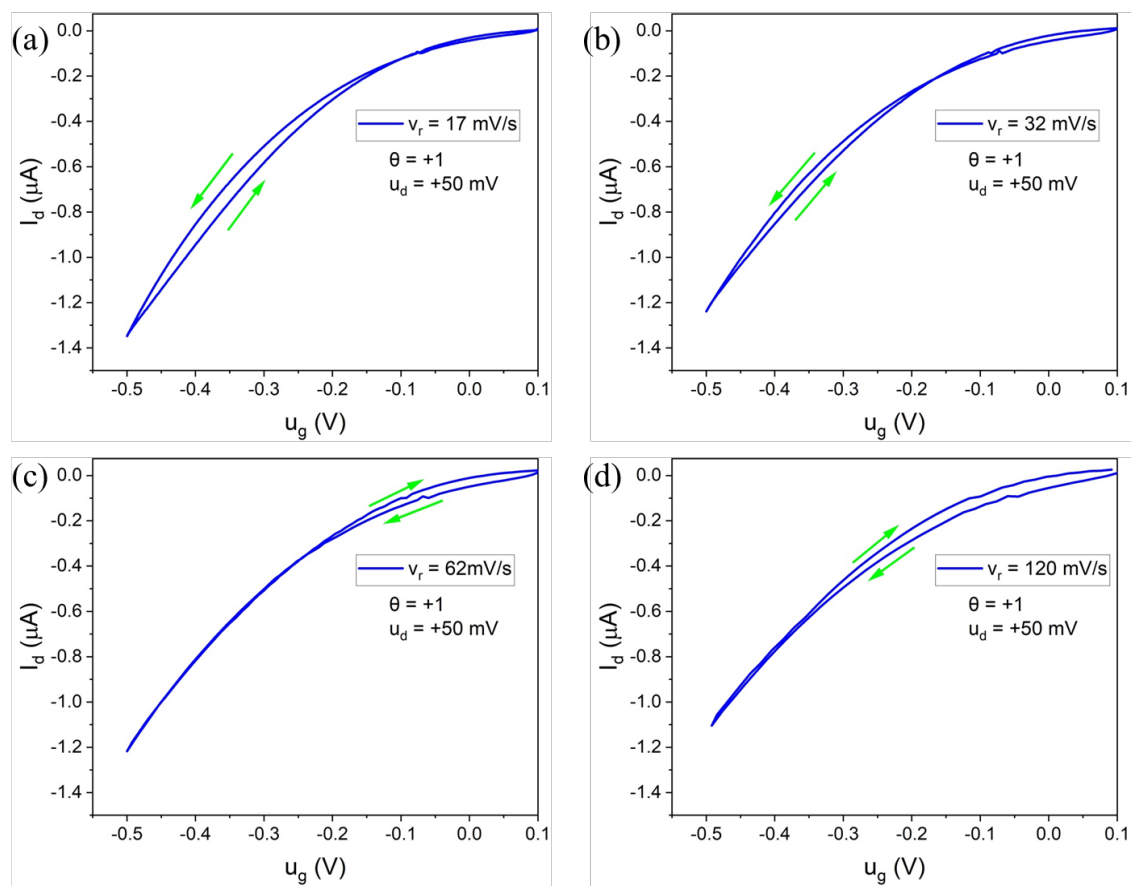

Fig. 11. Transfer curves of P3HT OECT for θ = +1 and $u_d$ = +50 mV at various scan rate.

For θ = -1, the increase in scan rate does not lead to a change in the hysteresis type (see Figure 12). We only observe an increase in the hysteresis strength and a decrease in the overall current level by increasing $v_r$, which is consistent with the theoretical model (Figure 6d).

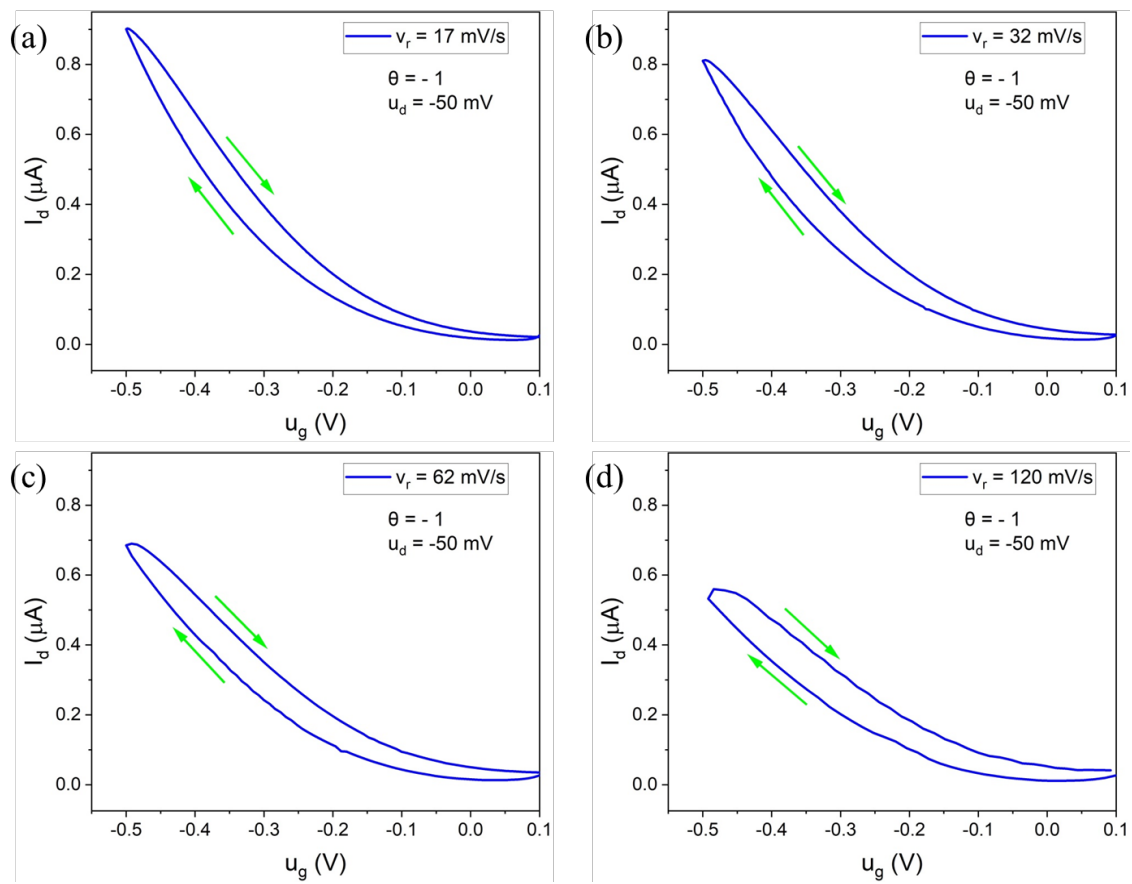

Fig. 12. Transfer curves of P3HT OECT for $\theta = -1$ and $u_d = -50$ mV at various scan rate.

## 6. Simulation of transients and the related hysteresis

The purpose of this Section is to bridge between the electrochemical model and experimental data using a device simulation that can include many details of the device structure. We start with a simulation that aligns with the chemical physics picture used by the electrochemical model. In terms of the 2D device simulations, [35] these translate to the following:

1) The electrolyte acts as an infinite supply of ions. We set its thickness to 200 μm.
2) The electrolyte presents no serial resistance. We set the ions' diffusion coefficient in the solution at least 6 orders of magnitudes above that of the diffusion in the 100 nm thick semiconductor.
3) There is no voltage drop at the gate electrode interface. The gate is ohmic to anions, and the anions density at the gate interface is set to its value in the solution. The contact acts as a reflecting mirror (non-reacting) for the cations.
4) The ionic double-layer capacitance ($C_{dl}$) at the source and drain electrodes is negligible. We set the effective double-layer thickness at 0.2 nm (i.e., areal capacitance is $\sim 10^{-5}$ Fcm$^{-2}$), and the contact overlap with the semiconductor ($L_c$) is kept well below 1 μm.

5) To avoid contact-limited injection, the source and drain contact barriers are set low enough such that the hole's current is not reduced due to the short overlap (100 nm). [74]

6) When the ions enter the semiconductor, they do not create a space charge that would impede their diffusion into the film. To ensure fast supply response of the compensating holes, we set their mobility at 5 cm$^2$V$^{-1}$s$^{-1}$ (For L = 50 μm and V$_{DS}$ = 0.1V it results in $\tau_e$ = 0.05 ms).

7) The mutual attraction, through Poisson equation, does not affect the dynamics. The dynamic response is evaluated at low ion densities (<10$^{18}$cm$^{-3}$) where the attraction is low.

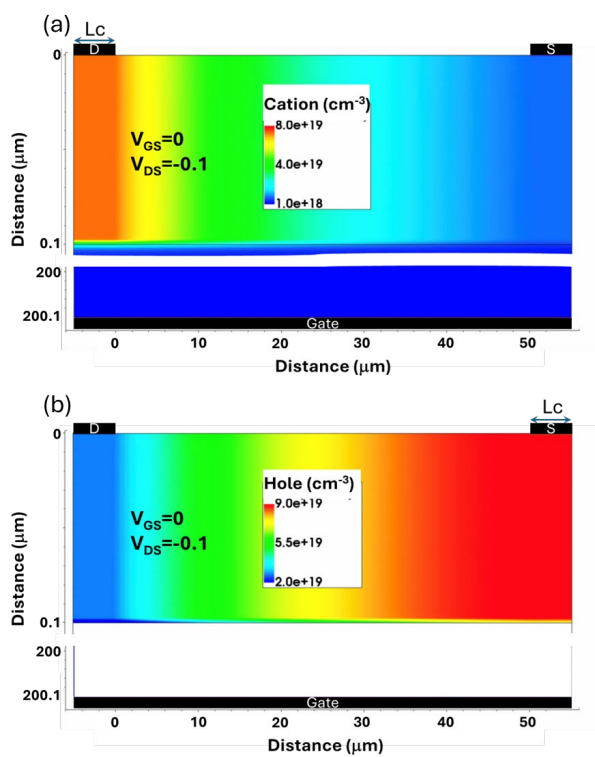

Fig. 13 The simulated device structure containing the parameters of a P-doped (10$^{20}$cm$^{-3}$) device. (a) Cation density (b) Hole density. V$_{GS}$=0V, V$_{DS}$=0.1V, W=L=50μm.

Fig. 13 shows the simulated device structure (w = L = 50 μm), on top of which are results from a simulation of a p-doped semiconductor (10$^{20}$ cm$^{-3}$) and an electrolyte ion concentration of 10$^{18}$ cm$^{-3}$ (~2 millimolar). Note that the choice of coordinates in the Fig. 13 implies that for $u_d < u_s$, the current is positive. Fig. 13a and Fig. 13b show the cation (a) and hole (b) density distribution under bias conditions of V$_{GS}$ = 0 and V$_{DS}$ = 0.1 V. The hole and cation distribution shown in Fig. 13 agree with previously reported simulation results. To follow the notation of the electrochemical model, the $I_{DS}$ is positive here.

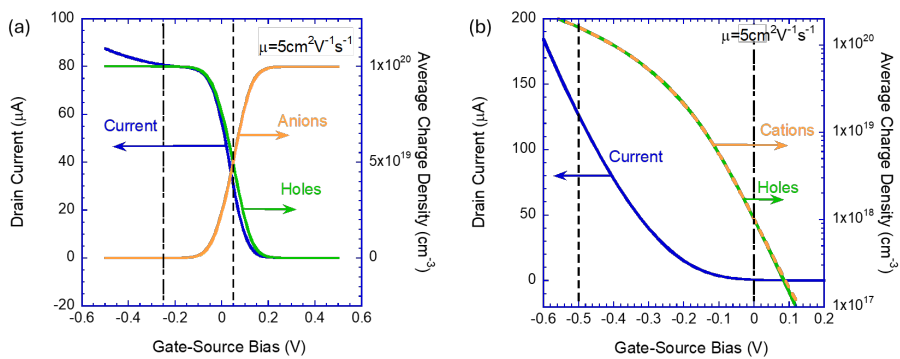

Fig. 14. Current (blue), hole (green), and ion (orange) density as a function of the gate voltage. (a) doped semiconductor. (b) undoped semiconductor. The dashed-dotted and the dashed black lines mark the voltage at which the low ion-density and high ion-density time-dependent responses are evaluated.

Fig. 14 shows the steady-state current-voltage (blue line) response of devices based on doped (a) and undoped (b) semiconductors. The average density of the hole (green line) and ions (orange line) in the semiconductor are plotted on the right axis. For the P-doped ($10^{20}$ cm$^{-3}$) device, we note that the hole density is a mirror image of the anion density following holes = doping - anions. For the undoped semiconductor (Fig. 14b), the holes, cations, and the current have a perfectly overlapping shape. Hence, we took the opportunity and plotted the charge densities on a log scale.

In the following sections, we will present transient responses in two regimes. The first is the regime where the ion density within the semiconductor is low ($<10^{18}$cm$^{-3}$), and we mark the relevant working points with dashed-dotted lines in Fig. 14. The second regime is for high ions density ($>10^{19}$cm$^{-3}$) with the relevant working point marked by a dashed line. Fig. 14 b shows that the current rises for gate-source bias below -0.2 V while the average hole density seems fixed. As reported in ref. [36], this is the signature of the thin hole channel generated by anions accumulation at the semiconductor interface.

**6.1 Transient response at low ion density**

To test for agreement between the semiconductor device simulations and the electrochemical model, we performed a transient response analysis at bias levels where the ion's density is low (dashed-dotted lines in Fig. 14). To ensure that the densities stay low throughout the response, we performed small signal analysis using a +30 mV step voltage at t=0. Fig. 15 shows the transient response for devices based on P-doped (top raw) and undoped (bottom raw) semiconductors. The left column shows the drain current ($\theta = 1$) and the middle one shows the source current ($\theta = -1$). We used hole mobility of $\mu_h = 5 cm^2V^{-1}s^{-1}$ ($\tau_e = 0.05ms$) and ion diffusion constant of D=10$^-$

$^6$cm$^2$s$^{-1}$ ($\tau_D = 0.1 ms$). Since $\tau_e < \tau_D$ the simulations of the doped device, Fig. 15a and Fig 15b correspond to Fig. 4f and Fig 4g, respectively. We note that the change in the shape of the responses between Fig. 15a and Fig. 15b agrees with the electrochemical model. Also, the stabilization time corresponds nicely to the ion's diffusion time ($\tau_d$). The difference between the source and drain currents during transients is best visualized in Fig. 15c. The fact that the source drives less current than the one extracted at the drain is the signature of the semiconductor being discharged following the positive bias step. The point where the source and drain current converge marks the end of the transient.[23,62]

Moving to the undoped device, Fig. 15d and Fig 15e correspond to Fig. 2f and Fig. 2g, respectively. We find that Fig. 15d does not agree with Fig. 2f, and when we place the source and drain currents next to each other (Fig. 15f), it seems as if no discharge takes place. This discrepancy suggests that the current of the undoped device is not just the hole current and that there is a significant contribution of displacement current ($\varepsilon$ dE/dt). With the aid of the simulation, we isolated the hole currents and plotted them as dashed lines in Fig. 15f. As they should, the hole currents are different enough to indicate a discharge following the step in the gate voltage. The electric field that is varying is the one between the source and drain even though V$_{ds}$ is constant. The non-uniformity of the ion's influx creates variations in the electric field while keeping the integral (i.e. V$_{ds}$) constant.

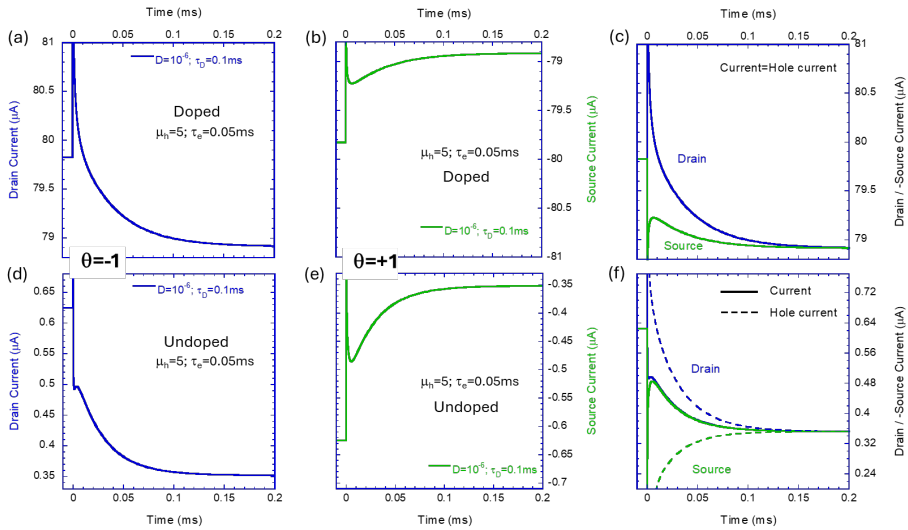

Fig. 15. Transient response of the drain current (V$_{ds}$=0.1V) to a small step in the gate voltage after being stabilized at the bias marked with dashed-dotted lines in Fig. 14. (a, b, c) P-doped (10$^{20}$cm$^{-3}$) semiconductor. (d, e, f) Undoped semiconductor. (a, d) drain current. (b, e) source current. (c, f) drain current and source current times -1.

Our next goal is to verify if making $\tau_e$ larger such that $\tau_e > \tau_d$ inverts the slope of the source current (θ = +1), which would indicate that the hysteresis loop reversed direction. As with Fig. 15, we find in Fig. 16 that the doped device (top raw) agrees with the electrochemical model while the undoped one (bottom raw) does not. Note that in Fig. 16b, the current flipped direction, and decreases towards the steady state value. The non-uniformity of the undoped device manifests itself with the source current (θ = +1, Fig. 16e) still rising towards the steady state. This poses a potential problem since the experimental data for the undoped P3HT device shows inverted hysteresis at low $V_{ds}$ bias. To address this issue, we simulate, in the following section, a device with a non-reacting gate electrode, as in the experiments.

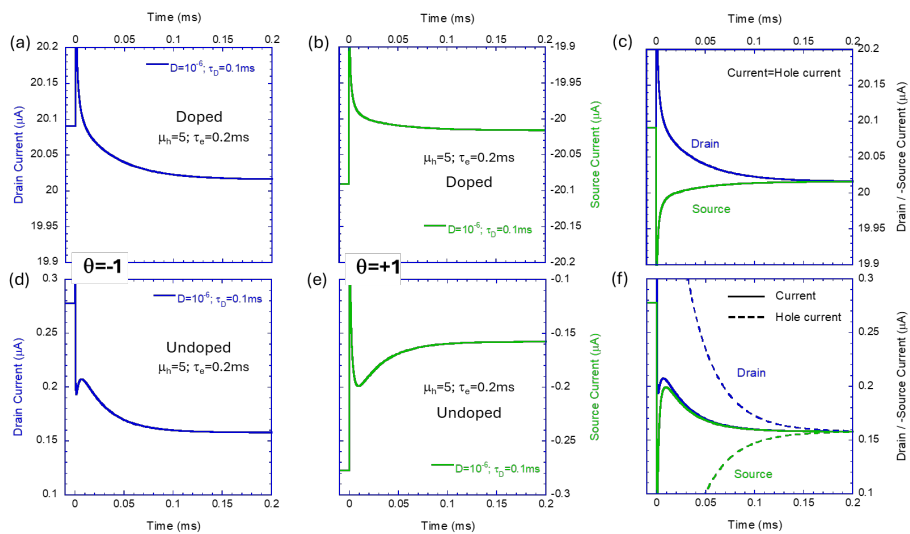

Fig. 16. Transient response of the drain current ($V_{ds}$ = 0.025 V) to a small step in the gate voltage after being stabilised at the bias marked with dashed-dotted lines in Fig. 13. (a, b, c) P-doped ($10^{20}$ cm$^{-3}$) semiconductor. (d, e, f) Undoped semiconductor. (a, d) drain current. (b, e) source current. (c, f) drain current and source current times -1.

**6.3 Hysteresis response and comparison to experiments**

This subsection addresses the transient response issue of predicting that the hysteresis loop for θ = +1 (source current) should not change direction for the undoped device at low $V_{ds}$ bias. This is important since the experimental data of the P3HT semiconductor (Fig. 10) show the hysteresis changing direction and low $V_{ds}$ bias (θ=+1, $V_{ds}$>0). Fig. 17a and Fig. 17b show the corresponding simulated hysteresis loops at several $V_{DS}$ values for the drain and source currents, respectively. We chose hole mobility (5 cm$^2$s$^{-1}$) and ion diffusion ($10^{-6}$ cm$^2$s$^{-1}$) so that at $V_{ds}$ = 50 mV, $\tau_e = \tau_d = 0.1\ ms$. The scan rate was such that when multiplied by the ion diffusion time ($\tau_d$) it equals 0.05 V. As expected from the simulated transient response, the hysteresis loop of the source current (θ = +1) does not change its direction at low drain-source bias.

Examining the experimental data (Fig. 10), we note that there is a sizable residual current at positive gate-source bias. This is the signature of residual doping, which is common to P3HT films, with some of the reported values exceeding $10^{16}$ cm$^{-3}$. To test if the residual doping could reduce the nonuniformity effects and recover the change in the hysteresis direction, we repeated the simulations with an added P-doping of $10^{16}$cm$^{-3}$. The results are presented in Fig. 17c and Fig. 17d for the drain and source currents, respectively. We note that, in agreement with the experimental data, the source ($\theta = +1$) currents' hysteresis loop direction changes at low $V_{ds}$ values (Fig. 17d). As stated in the experimental part, such a change in direction is not observed for the drain ($\theta = -1$) current (Fig. 17c).

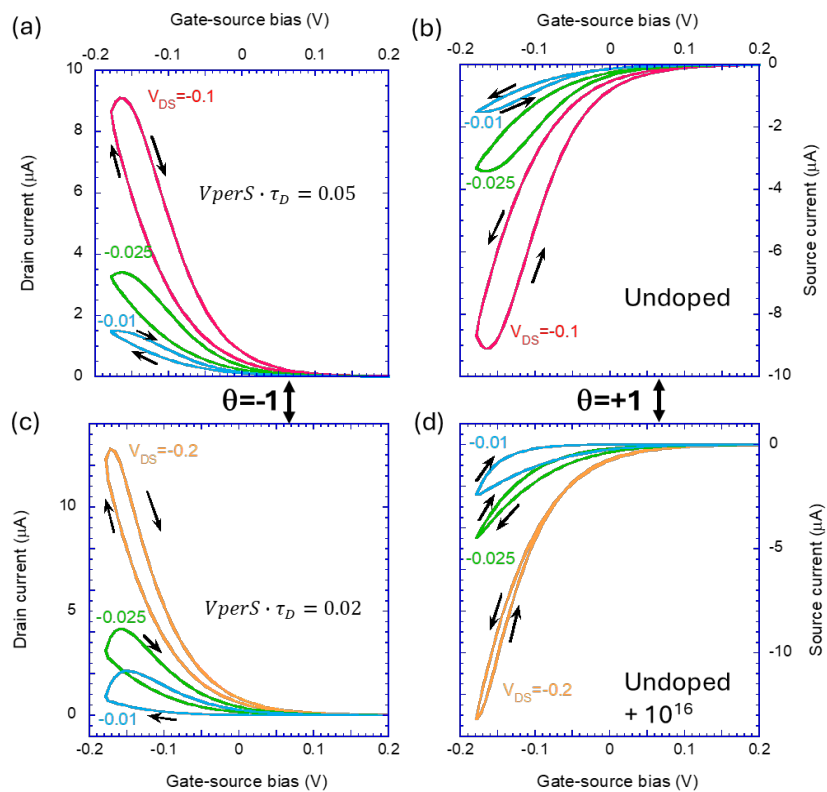

Fig. 17. The drain (left) and source (right) currents' hysteresis response to a voltage sweep in the gate voltage after being stabilised at the initial bias ($V_{gs}$=0.5V) and for several drain-source voltages (see figure). (a, b) The semiconductor is undoped, and the sweep speed is $0.05/\tau_d$. (c, d) The semiconductor is undoped but with residual P doping of $10^{16}$cm$^{-3}$, and the sweep speed is $0.02/\tau_d$.

## 6.2 Transient response at high ion density

Having bridged between the electrochemical model and the experimental results, we use the 2D semiconductor device simulations to explore an effect that is not part of the Bernards-Malliaras model. At the very heart of any electrochemical model, one can find the charge neutrality concept. Since (space) charge accumulation creates a repulsive electric field for added charge, it is impossible to accumulate significant charge in a given volume. For the steady-state solution, charge neutrality allows us to state that the density of holes equals (to an excellent approximation) the ion density. This simplifies the equations, as we only need to follow the ions. However, such a conclusion is not necessarily correct for the transient. During the transient, the charge neutrality principle dictates that ions (holes) cannot penetrate the film unless holes (ions) arrive to compensate for the ions' (holes') charge. Namely, in the general case, the dynamics depend upon both $\tau_e$ and $\tau_d$. Naturally, the cross dependence, which is driven by the Poisson equation, is a function of the charge (ion) density. In the previous sections, we minimized this cross-dependence by limiting the gate-source bias to a range where the ion density is below $10^{18}$ cm$^{-3}$. In the following figure we will show results for the range in which the ion density is above $10^{19}$ cm$^{-3}$.

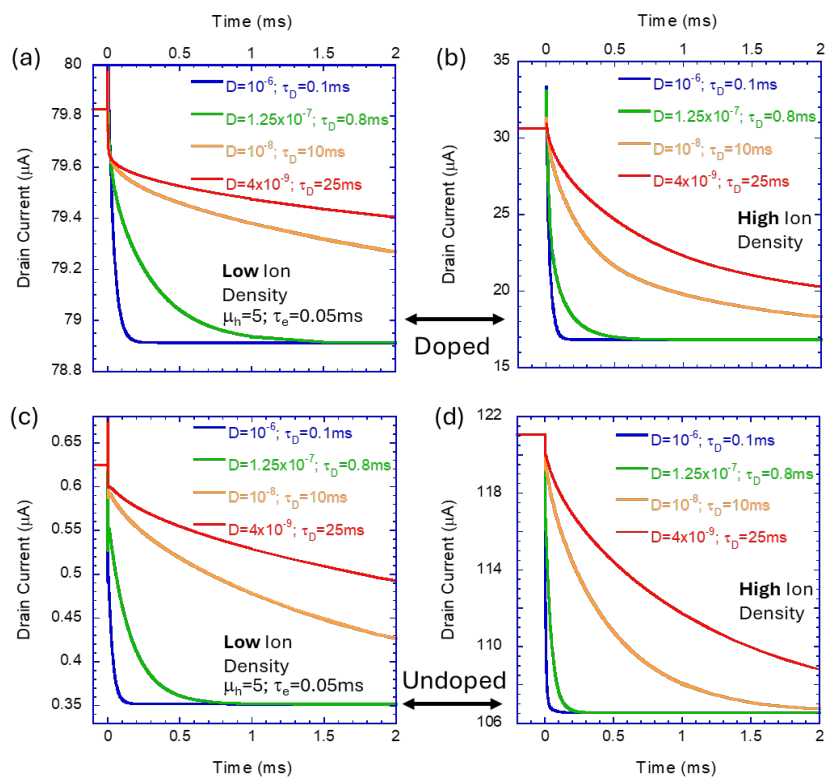

Fig. 18. Transient response of the drain current to a small step in the gate voltage after being stabilized at the initial bias ($V_{DS}$=0.1V). (a, b) Doped semiconductor. (c, d) undoped semiconductor. (a, c) Low ion density (dashed-dotted lines in Fig. 13). (b, d) High ion density (dashed lines in Fig. 13). For the low ion density, the stabilization time is similar to the ion diffusion time ($\tau_d$) and for the high ion density it is much faster.

In Fig. 18 we look at the effect of the hole transit time being much shorter than the ion's diffusion time ($\tau_e \ll \tau_d$). We examine only the drain current ($\theta$=-1), with the left column for the low-density case and the right one for the high-density. For comparison, we added results from longer diffusion times. Both Fig. 18a (doped) and Fig. 18c (undoped) shows that the stabilisation time constant equals the ion diffusion time and the use of $exp\left(-\frac{t}{\tau_d}\right)$, is justified. The high-density results, which appear in the right column, are very different. The stabilization time constant still depends on the diffusion time, but it is now much shorter. Notably, the orange line representing the diffusion time of 10ms exhibits a stabilization time of about 2ms. Apparently, the fast holes expedite the transport of the ions.

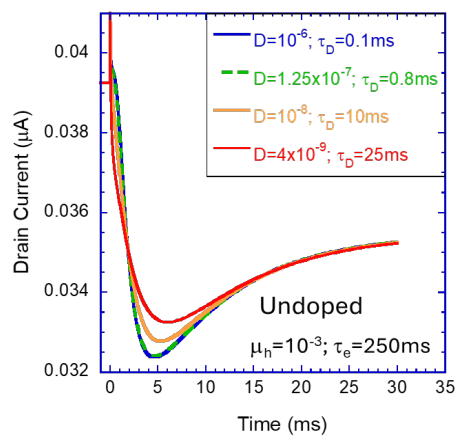

Fig. 19. The drain current's transient response to a small step in the gate voltage after being stabilized at the initial bias ($V_{DS}$=0.1V). The hole transit time is much longer than the ion diffusion time ($\mu_h$=10$^{-3}$cm$^2$V$^{-1}$s$^{-1}$, $\tau_d$=250ms). The semiconductor is undoped, and the response is at high ion density (dashed line in Fig. 13).

While Fig. 18 showed that holes entering fast would expedite the ions, we use Fig. 19 to demonstrate that slow holes delay the ions' response. In Fig. 19 the hole mobility is reduced to 10$^{-3}$cm$^2$V$^{-1}$s$^{-1}$ resulting in $\tau_e$=250 ms. All the lines, starting from $\tau_d$=0.1ms (blue line) and moving all the way to $\tau_D$=25ms (red line), converge together towards the steady state at longer than 30ms. Again, it shows that at high ion density the response time depends on both the hole transit time ($\tau_e$) and the ion's diffusion time ($\tau_d$).

## Conclusion

We developed a general formulation for a 2D transmission line description of an OECT. The general model considers the horizontal transport of electronic carriers and vertical diffusion of compensating ions, without space charge effects. Then we provided an analytical time dependent model under assumptions of homogeneous charge distribution. This model is equivalent to Bernards-Malliaras description based on a variable ion concentration along the channel distance. However we have extended the BM model by including the diffusion effect, so that two time constants have been identified, the horizontal hole transport $\tau_e$ and the vertical ion diffusion $\tau_d$, depending on transport coefficients, morphology, and horizontal field. This model enables a basic classification of the transient response to a step voltage in terms of the sign of the current and the sign of the step. Based on this elementary responses we found the possible types of hysteresis behaviour in transfer curves. The experimental results validate the theoretical model, demonstrating the influence of various parameters such as drain bias, scan rate, and polarity on the hysteresis behavior observed in OECT transfer curves. The analysis of realistic simulation results shows very good agreement with the simple model, hence the model produces a robust elementary classification of the dynamic responses of OECT that are mainly governed by ion diffusion and hole transport issues. However, additional effects occur in realistic simulation, since the carrier density affects the effective transport parameters and change the results expected in single carrier parameter pictures.

These findings contribute to a deeper understanding of the underlying mechanisms governing the operation of organic electrochemical transistors.

## Acknowledgments

The work of Juan Bisquert was funded by the European Research Council (ERC) via Horizon Europe Advanced Grant, grant agreement nº 101097688 ("PeroSpiker"). Baurzhan Ilyassov thanks the Science Committee of the Ministry of Science and Higher Education of the Republic of Kazakhstan for financial support under the project Grant No. AP13067629. Nir Tessler acknowledges the support by the Ministry of Innovation, Science and Technology Israel, the M-ERANET grant PHANTASTIC Call 2021.

## Associated content

Data Availability Statement

The data presented here can be accessed at https://doi.org/10.5281/zenodo.10925192 (Zenodo) under the license CC-BY-4.0 (Creative Commons Attribution-ShareAlike 4.0 International).